\documentstyle[12pt] {article}
\begin {document}
\parindent=15pt
\begin{center}
\vskip 1.5 truecm
{\bf CROSS SECTIONS, PROBABILITIES, MULTIPLICITIES AND SPECTRA OF 
SECONDARIES IN HIGH ENERGY HEAVY ION INTERACTIONS}\\
\vspace{.5cm}
J.Dias de Deus and Yu.M.Shabelski$^*$  \\
\vspace{.5cm}
CENTRA, Instituto Superior Tecnico, 1049-001 Lisboa, Portugal \\
\vspace{.5cm}

\end{center}
\vspace{1cm}
\begin{abstract}
We present a short review of theoretical results (mainly for
experimentalists) published in many different papers. The formulae are
presented for the different integrated cross sections, the number of
interacting nucleons, multiplicities of secondaries, the dispersions
of multiplicity distributions. Two possible tests for the search of
Quark-Gluon Plasma formation are discussed. CERN SPS data for production 
of secondaris in central Pb+Pb collisions are compared with Quark-Gluon
String Model predictions. 

\end{abstract}
\vspace{3cm}

Permanent address: \\
Petersburg Nuclear Physics Institute, \\
Gatchina, St.Petersburg 188350 Russia \\

\newpage

\section{Introduction}
At present  time, different Monte Carlo codes are used, as a rule,
for the analyses of existing heavy ion experiments and for planning
the future ones. However, sometimes it is not clear enough, what
physics was used in some Monte Carlo code, which processes were
accounted for, and so on. In the present paper we give a review
of the simplest theoretical predictions which can be used for
different estimates, not involving complicated calculations.

Formally we consider the high energy nucleus-nucleus collision as a
superposition of the independent nucleon-nucleon interactions.
However, really the main part of the discussed results is based
practically only on geometry, and does not depend on the model of
interaction.

\section{Cross sections}

Let us consider the collision of nucleus $A$ with nucleus $B$, assuming
that they are not very light. The simplest expression for the total
inelastic $\sigma^{in}_{AB}$, or secondary production cross section
$\sigma^{prod}_{AB}$ comes from geometry:
\begin{equation}
\sigma^{in}_{AB} \approx \sigma^{prod}_{AB} = \pi (R_A + R_B)^2 \;.
\end{equation}
The difference between $\sigma^{in}_{AB}$ and $\sigma^{prod}_{AB}$, which
corresponds to all processes of disintegration or exitation of one or both 
nuclei without secondary production, is negligibly small at high energies 
and heavy incident nuclei. In the case of heavy ion collisions the total 
elastic cross section is only slightly smaller than $\sigma^{in}_{AB}$ 
\cite{KS}, $\sigma^{el}_{AB} \approx \sigma^{in}_{AB}$, so
\begin{equation}
\sigma^{tot}_{AB} = \sigma^{el}_{AB} + \sigma^{in}_{AB} =
2\pi (R_A + R_B)^2 \;.
\end{equation}

More accurate is Bradt-Peters expression \cite{BP}
\begin{equation}
\sigma^{in}_{AB} = \pi R_0^2 (A^{1/3} + B^{1/3} - c)^2 \;,
\end{equation}
which accounts for the possibility of very peripheral collisions without
inelastic interactions. The values of parameters $R_0$ and $c$ are
slightly different for light and heavy nuclei \cite{KS}. The comparison
of this expression with the experimental data was presented in
\cite{Sh1}.

More detailed results for different cross sections of heavy ion 
interactions can be obtained with the help of multiple scattering 
(Glauber) theory, see \cite{BS} for details. This theory can be used at 
all energies starting from several hundred MeV per nucleon. At high 
energies (say, more than 10 GeV per nucleon) in principle the inelastic 
screening \cite{Gr,KM} should be accounted for, however in the case of 
heavy ion cross sections these effects are numerically very small.

The amplitude of $A-B$ elastic scattering with momentum transfer $q$ can 
be written in the frame where $B$-nucleus is a fixed target as
\begin{equation}
F^{el}_{AB}(q) = \frac{ik}{2\pi} \int d^2b e^{iqb} [1 - S_{AB}(b)] \;,
\end{equation}
where $k$ is the incident momentum of one nucleon in $A$-nucleus,
$b$ - impact parameter, and
\begin{equation}
S_{AB}(b) = \langle A \vert \langle B \vert
\prod_{\stackrel{i \in A}{j \in B}} [1 - \Gamma_{NN}(b + u_i - s_j)] \;
\vert B \rangle \vert A \rangle \;,
\end{equation}
\begin{equation}
\Gamma_{NN}(b + u_i - s_j)] = \frac{1}{2 \pi ik} \int d^2q
e^{-iq(b + u_i - s_j)} f_{NN}(q) \;,
\end{equation}
where $u_i$ and $s_i$ are the transverse coordinates of nucleons.
Contrary to the case of hadron-nucleus interaction, we can not integrate
Eq. (5) analitically even after standard assumption that nuclear density 
is a product of one-nucleon densities. 

To make the problem manageable, one can pick up a certain fraction only of 
various contributions in the expansion of the product in Eq. (5), 
distinguished by large combinatorial factors. The leading graphs 
correspond to the so-called optical approximation \cite{CM}, which sums up 
the contributions with scattering not more than once for each nucleon, 
i.e. takes into account the products of amplitudes $\Gamma_{NN}$ in Eq.(5) 
with all indices $i$, $j$ being different. It corresponds to the summation 
of diagrams shown in Fig. 1 and it contain the largest combinatorial 
factor $A(A-1)...(A-n+1)B(B-1)...(B-n+1)$ in the case of $n$-fold 
interaction. In all another series the combinatorial factors are 
significantly smaller, however these series have different signs, so due 
to the large cancelation some of them can give significant contribution to 
the final expression. To avoid crowding of lines we have only shown in 
Fig. 1 the nucleon participants from the nucleus $A$ (upper dots) and $B$ 
(lower dots) with the links standing for interacting amplitudes, and we 
do not plot the nucleon-spectators.

   In the language of Eq. (5) the optical approximation can be
reproduced, making the averaging $\langle A \vert ... \vert A \rangle$
and $\langle B \vert ... \vert B \rangle$ inside the product
$\prod_{i,j}$ in Eq.(5) instead of averaging the product as a whole:
\begin{equation}
S_{AB}^{opt}(b) = \prod_{i,j} \langle A \vert \langle B \vert
[1 - \Gamma_{NN}(b + u_i - s_j)] \; \vert B \rangle \vert A \rangle \;,
\end{equation}

Using the standard assumptions of the multiple scattering theory one
obtains
\begin{equation}
S_{AB}^{opt}(b) = \left [1 - \frac1A T_{opt}(b) \right ]^A \approx
exp [- T_{opt}(b)]
\end{equation}
with
\begin{equation}
T_{opt}(b) = \sigma/2 \int d^2b_1 T_A(b_1-b) T_B(b_1)
\end{equation}
and
\begin{equation}
T_A(b) = A \int^{\infty}_{-\infty} dz \rho (b=\sqrt{r^2 - z^2}, z)
\end{equation}
where $\rho(r)$ is the one-particle nuclear density,
$\int \rho(r) d^3r = 1$.

   In the Reggeon language the optical approximation corresponds to the
accounting only one pole (nuclear ground state) in the both $M_A^2$ and
$M_B^2$ complex planes instead of all intermediate states, see \cite{BS}.

However, numerical calculations \cite{FV} (see also \cite{Sh2} for the
case of light nuclei collision) demonstrate that the optical
approximation is not accurate enough even for the total cross sections.
The difference with the data amounts $\sim$ 10-15\% for
$\sigma_{AB}^{tot}$ and is much larger for differential cross sections.

The more explicit rigid target \cite{Alk,VT,CGS,PR} approximation can be
reproduced, making the averaging only
$\langle B \vert ... \vert B \rangle$ inside the product in Eq. (5) :
\begin{equation}
S_{AB}^{r.g.}(b) = \langle A \vert \{ \prod_{i,j} \vert \langle B \vert
[1 - \Gamma_{NN}(b + u_i - s_j)] \; \vert B \rangle \}
\vert A \rangle = [T_{r.g.}(b)]^A \;,
\end{equation}
\begin{equation}
T_{r.g.}(b) = \frac1A \int d^2b_1 T_A(b_1 - b)
exp \left [ - \frac{\sigma}{2} T_B(b_1) \right ] \;,
\end{equation}

This approach corresponds to the diagrams of Fig. 2, where each nucleon
from the nucleus $A$ can interact several times, but all interacting
nucleons from $B$ are still different. Due to the obvious asymmetry in
contributions of the two nuclei such approach can be justified
theoretically in the limit $A/B \ll 1$, say for $C-Pb$, or $S-U$
collisions, however sometimes it can be used for the case of heavy ion
collisions with equal atomic weights, see next Sect.

Further corrections to the elastic amplitude have been considered in
Refs. \cite{AC,AK,Pak,Bra,BK}, however these results are rather
complicated for practical use.

Another possibility is the direct calculation of Eq. (5) using Monte
Carlo simulation \cite{KS,ZUS}.

Let us present now the list of several integrated cross sections.
The total cross section of  $A-B$ collision is given by the optical 
theorem
\begin{equation}
\sigma^{tot}_{AB} = 2 \int d^2b [1 - S_{AB}(b)] \;,
\end{equation}
where the function $S_{AB}(b)$ can be taken, say, from Eq. (8) or
Eq. (11). The integrated cross section of elastic $A-B$ scattering is
\begin{equation}
\sigma^{el}_{AB} =  \int d^2b [1 - S_{AB}(b)]^2 \;.
\end{equation}
The sum of cross sections of all diffractive processes without
production of secondaries, i.e. the elastic and quasielastic scattering
can be calculated with the help of close approximation
$\sum_{A'} \vert A' \rangle \langle A' \vert = 1$ similarly to the
case of hadron-nucleus scattering \cite{FG}
\begin{equation}
\sigma^{scat}_{AB} =  \int d^2b \{1 - 2S_{AB}(b) + [I_{AB}(b)]^A \} \;,
\end{equation}
where in the optical approximation
\begin{equation}
I_{AB}^{opt}(b) = 1 - \frac1A R_{opt}(b)  \;,
\end{equation}
\begin{equation}
R_{opt}(b) = \sigma^{inel} \int d^2b_1 T_A(b_1-b) T_B(b_1)
\end{equation}
and $\sigma^{inel}$ is the $NN$ inelastic cross section.
In the rigid target approximation
\begin{equation}
I_{AB}^{r.t.}(b) = \frac1A \int d^2b_1 T_A(b_1-b)
exp[-\sigma^{inel}T_B(b_1)]
\end{equation}

The correspondent expression for the secondary production cross section
reads
\begin{equation}
\sigma^{prod}_{AB} = \sigma^{tot}_{AB} - \sigma^{scat}_{AB} =
\int d^2b \{1 - [I_{AB}]^A \} \;.
\end{equation}

It is also possible \cite{ST} to obtain the integrated stripping cross 
sections of collisions among light nuclei. This can be done by considering 
the different intermediate states \cite{AGK,Shab,Sh2} of all Glauber 
diagrams for elastic AB amplitude. The resulting expressions have the 
simple form of a combination of several total inelastic cross sections. 
For example, the cross section for stripping one nucleon from nucleus A in 
A-B collision, $\sigma^{(1)}_{AB}$, is written as 
\begin{equation}
\sigma^{(1)}_{AB} = A(\sigma^{inel}_{AB} - \sigma^{inel}_{(A-1)B}) \;,
\end{equation}
where the density distribution of the nucleus with atomic weight A-1
should be slightly  "corrected" \cite{ST}

\section{Distributions on the number of interacting nucleons}

Let us consider the events with secondary hadron production in
nuclei A and B minimum bias collisions. In this case the average number
of inelastically interacting nucleons of a nucleus A is equal \cite{BBC}
to
\begin{equation}
<N_A>_{m.b.} = \frac{A \sigma^{prod}_{NB}}{\sigma^{prod}_{AB}} \;.
\end{equation}

If both nuclei, A and B are heavy enough, the production cross sections
of nucleon-nucleus and nucleus-nucleus collisions can be written as
\begin{equation}
\sigma^{prod}_{NB} = \pi R_B^2 \;, \;\;
\sigma^{prod}_{AB} = \pi (R_A + R_B)^2 \;, \;\;
\end{equation}
and accounting for Eq. (1) we obtain that in the case of
equal nuclei, A = B, for minimum bias (m.b.) events which are averaged 
over impact parameter
\begin{equation}
<N_A>_{m.b.} = A/4 = <N_A>_{c}/4 \;.
\end{equation}
So in the case of minimum bias events the average number of interacting
nucleons should be four times smaller than in the case of central
collisions \cite{PSh}, where (neglecting by small corrections which will
be discussed below) $<N_A>_c \approx A$.

The average number of inelastic $NN$ interactions in minimum bias $AB$ 
collisions with any secondary production is equal \cite{Sh1,PR} to
\begin{equation}
<\nu_{AB}>_{m.b.} = A B \sigma^{inel}_{NN}/\sigma^{prod}_{AB} \;.
\end{equation}

Let us note that the optical approximation cannot be used \cite{Sh1}
for the calculations of the averaged numbers of interacting nucleons 
and the distribution on this number. The rigid target approximation gives 
here reasonable results. In particular, for the distribution over the 
number of inelastically interacting nucleons $N_A$ of A nucleus in $AB$ 
interaction we have
\cite{CGS}
\begin{equation}
V(N_A) = \frac{1}{\sigma^{prod}_{AB}} \frac{A!}{(A-N_A)! N_A!}
\int d^2 b [I^{r.t.}_{AB}(b)]^{A-N_A} [1-I^{r.t.}_{AB}(b)]^{N_A} \;,
\end{equation}

The theoretical predictions for the distributions on the electrical
charge of non-interacting nucleons based on Eq. (25) were compared with 
the data in \cite{CGS} and the agreement was quite reasonable, see 
Fig. 3 taken from \cite{CGS}.

All the presented results are related to the case when only the events
with secondary production are registrated. In the case when the events
with nuclear disintegration without secondary production are also
registrated, one should change $\sigma^{inel}_{NN}$ by
$\sigma^{tot}_{NN}$ and  $\sigma^{prod}_{AB}$ by $\sigma^{inel}_{AB}$.

Eq. (25) is written for minimum bias events. In the case of events
for some interval of impact parameter $b$ values, the integration
in Eq. (25) should be fulfilled by this interval,
$b_{min} < b < b_{max}$. In particular, in the case of central
collisions the integration should be performed with the condition
$b \leq b_0$, and $b_0 \ll R_A$. In the case of the collisions of equal
heavy ions the value of $<\!N_{A}\!>$ decreases with increase of the
impact parameter. As even at small $b \neq 0$ some regions of
colliding ions are not overlapping, only very small fraction of events,
not larger than 0.5-0.7\% of all minimum bias sample, can be considered as
the central interactions.

In the case of asymmetrical ion collisions, say $S-U$, all nucleons of
light nucleus at small impact parameters go throw the regions of
relatively high nuclear matter density of heavy nucleus, so
practically all these nucleons interact inelastically. For the case of
$S-U$ interactions at $\sqrt{s_{NN}}$ = 20 GeV all events with
$b < 2\div 3$ fm can be considered as the central ones \cite{PSh}.

The predictions for the distributions on the number of inelastically
interac\-ting nucleons at different impact parameters can be found in
\cite{PSh}.

\section{Multiplicities of secondaries, inclusive densities and
tests for QGP formation}

Let us now look at the inclusive density of the produced secondaries,
\begin{equation}
\rho_{AB}(x) = E \frac{d^3\sigma^{AB}}{d^3p} \;.
\end{equation}
In the case of asymptotically high energies we predict\footnote{We do not
account here the additional shadowing due to percolation \cite{DDU} or
multiple pomeron interactions \cite{CKT}.} for $x_F \to 0$ \cite{Sh1}
\begin{equation}
\rho_{AB}(x)/\rho_{NN}(x) = <\nu_{AB}>
\end{equation}
and for secondary multiplicities
\begin{equation}
<n_{AB}>/<n_{NN}> = <\nu_{AB}>
\end{equation}

However at the realistic energies both right-hand side ratios (27) and 
(28) are significantly smaller than the predicted values due to the 
energy conservation corrections. For example, if one nucleon
of the projectile nucleus interact with several nucleons of the target
nucleus, the total energy of all secondaries should be equal to the
initial energy, so the effective energy of every $NN$ interaction
is smaller than the initial energy. These effects are more important in
the both $A$ and $B$ fragmentation regions.

The corrections to Eqs. (27) and (28) decrease with the growth of the
initial energy. Some predictions for the multiplicity of secondaries
with accounting the division of energy between several $NN$ interactions
can be found in \cite{Sh3}.

The relations between average multiplicities and inclusive densities
of secondaries produced in heavy ion collisions can be used for search
of the Quark-Gluon Plasma (QGP) formation. One of them is the relation
between average multiplicity and inclusive density in the central
region. In the case of $NN$ collisions for any $z = n_{-NN}/<n_{-NN}>)$ 
the ratio \footnote{The considering ratio is rather
similar to KNO one, however instead of cross sections we use inclusive
densities.} 
\begin{equation}
R(z) = \rho_{NN}(x)/<\rho_{NN}(x)> \vert_{x \sim 0}
\end{equation}
is a non-trivial function of $z$ \cite{RSh}. The reason is again in energy 
conservation. Let us compare two events, first with $n_- = <n_->$ and 
second with $n_- = 2<n_->$. Evidently, if the shapes of the inclusive 
spectra in these events will be the same, the energy of all secondaries in 
the second event will be two times larger than in the first one. So the 
shapes should be different, in the second event we will find smaller 
number of fast secondaries and larger number of slow ones. The 
experimental data \cite{Aln} as well as the model calculations \cite{RSh} 
show that the ratio $R(z) = \rho_{NN}(x)/<\rho_{NN}(x)> \vert_{x \sim 0}$ 
is about 3.5 at $z = 2.5$.

However, in the case of heavy ion collisions with independent
nucleon-nucleon interactions we can not expect any difference of Eq. (29)
from the linear function, because every independent nucleon-nucleon
collision gives the same contribution to $\rho_(x)$ as well as to
multiplicity \footnote{The difference in heavy ion and $NN$ interactions 
here is connected both with experimental conditions and with 
confinement. In heavy ion collisions we usually registrate only the 
interacting nucleons and neglect the non-interacting ones. Due to that 
the energy of secondaries can change from event to event. In the case of 
$NN$ interaction the spectator quarks and gluons will be involved into
the interaction due to confinement forces, and the energy of all 
secondaries is exactly equal to the initial energy.}. The possible 
violation of such linear dependence should be considered as an evidence 
for some collective interaction (possibly, QGP formation). The analysis of 
the data at energy about 4 GeV per nucleon \cite{BKS} shows good agreement
with linear dependence, and the similar analysis of the data at higher
energies seems to be very interesting. Such analysis can be provided 
in some kinematical domain.

The second test for QGP formation is connected with analysis of the
multiplicities of different secondaries produced in the central and
minimum bias $A-B$ collisions \cite{PSh}. In the case of independent
nucleon-nucleon collisions the multiplicity of secondaries produced in
the central region should be proportional to the number of
interacting nucleons of projectile nucleus, that is four times larger
than in minimum bias events, see Eq. (23). It should depends also on
the average number, $<\! \nu_{NB} \!>$, of inelastic interactions of
every projectile nucleon with the target nucleus. As was shown in
\cite{Sh4}, the average number of interactions in the case of central
nucleon-nucleus collisions, $<\! \nu \!>_c$, is approximately 1.5 times
larger than in the case of minimum bias nucleon-nucleus collisions,
$<\! \nu \!>_{m.b.}$. After accounting several another corrections
\cite{PSh} one obtain
\begin{equation}
<\! n \!>_c \sim 4.5 <\! n \!>_{m.b.} \;.
\end {equation}

In the conventional approach considered here, we obtain the prediction
of Eq. (30) for any sort of secondaries including pions, kaons,
$J/\psi$, Drell-Yan pairs, direct photons, etc. Let us imagine that the
QGP formation is possible (or dominate) only at comparatively small
impact parameters, i.e. in the central interactions. In this case Eq.
(30) can be strongly violated, say, for direct photons and, possibly,
for light mass Drell-Yan pairs, due to the additional contribution to
their multiplicity in the central events via the thermal mechanism. At
the same time, Eq. (30) can be valid, say, for high mass Drell-Yan pairs 
and for pions, if the most part of them is produced at the late stage of 
the process, after decay of the plasma state. So the experimental 
confirmation of Eq. (30), say, for pions and its violation for the 
particles which can be emitted from the plasma state should be considered 
as a signal for QGP formation. Of course, the effects of final state 
interactions, etc. should be accounted for in such test.

\section{Multiplicity distributions and dispersions}

It was shown in Ref. \cite{DPS} that the main contribution to the
dispersion of multiplicity distribution in the case of heavy ion
collisions comes from the dispersion in the number of nucleon-nucleon
interactions. The last number is in strong correlation
with the value of impact parameter.

For the normalized dispersion $D/<\! n \!>$, where
$D^2 = <\! n^2 \!> - <\! n \!>^2$
we have \cite{DPS}
\begin{equation}
\frac{D^2}{<\! n \!>^2} =
\frac{<\nu_{AB}^2> - <\nu_{AB}>^2}{<\nu_{AB}>^2}
+ \frac{1}{<\nu_{AB}>} \frac{d^2}{\overline{n}^2} \;,
\end {equation}
where 
\begin{equation}
<\nu_{AB}> = <\! N_A \!> \cdot <\nu_{NB}> 
\end {equation}
is the average number of nucleon-nucleon interactions in nucleus-nucleus 
collision, $\overline{n}$ and $d$ are the average multiplicity and the 
dispersion in one nucleon-nucleon collision.

In the case of heavy ion collisions $<\nu_{AB}> \sim 10^2 - 10^3$, so,
as a rule, the second term in the right hand side of Eq. (31) becomes 
negligible \cite{DPS}, and the first term, which is the relative 
dispersion in the number of nucleon-nucleon interactions, dominates. In 
the case of minimum bias A-B interaction the last dispersion is 
comparatively large due to large dispersion in the distribution on $N_A$
($\nu_{AB}$). So in the case of some trigger (say, $J/\psi$ 
production) without fixing of the impact parameter, the multiplicity of 
all another secondaries can change significantly in comparison with its 
average value. 

In the case of some narrow region of impact parameters we have the 
opposite situation. Dispersion in the distribution on $N_A$ 
($\nu_{AB}$) is very small \cite{PSh}, especially in the case of 
central collisions. The dispersion in the number of inelastic 
interactions of one projectile nucleon with the target nucleus, 
$\nu_{NB}$, should be the same or slightly smaller in comparison with 
the minimum bias case. So the dispersion in the multiplicity of the 
produced secondaries can not be large. 

If fluctuations in the effective number of interactions are negligible, 
the second term in the right hand side of Eq. (31) should dominate and the 
dispersion of the produced secondaries will be suppressed by the factor
$<\nu_{AB}>$ in comparison with the case of $NN$ collisions. 

In the case of small fluctuations in $\nu_{AB}$ one can see that any 
trigger can not change significantly the average multiplicity of 
secondaries in the central heavy ion collisions, even if this trigger 
strongly influents on the multiplicity in the nucleon-nucleon 
interaction.

Some numerical example of very narrow secondary multiplicity
distribution in central heavy ion collisions can be found in 
\cite{Sh3}.

Now let us consider some rare event $C$ in high energy heavy ion 
collisions. It can be Drell-Yan heavy mass pair, $J/\Psi$, $\Upsilon$, 
$W$ or $Z$ production. The normalized multiplicity distribution of the 
produced secondaries in the events associated to such rare events, 
$P_C(n)$, should be not the same as the distribution in the mimimum bias 
events, $P_{m.b.}(n)$, with $\sum P{m.b.}(n) = \sum P_C(n) = 1$. The 
reason is quite clear. The rare events can occur independently in every 
$NN$ collision, so in the case of heavy ion interactions the probability 
of rare event is proportional to $\nu_{AB}$, Eq.~(32). So \cite{DPS1}    
\begin{equation}
P_C(n) = \frac{\nu_{AB}}{<\nu_{AB}>} P(n) 
\end {equation}

The experimental data at high energies are in agreement with the 
prediction of Eq.~(33). An example of such agreement is presented in
Fig. 4, taken from \cite{DPS1}. The similar relation can be written for 
the distributions on the transverse energy, $E_T$ \cite{DPS2,DP}, etc., 
and they are in agreement with existing high energy data 
\cite{DPS1,DPS2}.

\section{Spectra of secondaries}

As we discussed, the energy conservation effects violate asymptotical
ratios (27) and (28). The possible way to account for these effects was 
suggested in \cite{Sh3,Sh4}. The inclusive spectrum of secondaries 
produced in nucleus A - nucleus B collision can bew written as
\begin{eqnarray}  
\frac1{\sigma^{prod}_{AB}} \frac{d \sigma (AB \to hX)}{d y} & = &
\theta (y) R^h_A(y) <N_A> \frac1{\sigma^{prod}_{NB}} 
\frac{d \sigma (NB \to hX)}{d y} +  \\ \nonumber 
& + &\theta (-y) R^h_B(-y) <N_B> 
\frac1{\sigma^{prod}_{NA}} \frac{d \sigma (NA \to hX)}{d y} \;,
\end{eqnarray}
where $y$ is the rapidity of secondary $h$ in c.m. frame and the 
functions $R^h_{A,B}(y)$ account for the energy conservation effects.
   Here we connect the inclusive spectra of secondaries in the heavy ion 
collisions with the spectra in the nucleon-nucleus interactions. To 
calculate the last ones as well as the functions $R^h_{A,B}(y)$ we use one 
of the variants of the dual topological unitarization approach - the 
quark-gluon string model (QGSM) \cite{KTM}. This model has been used 
successfully to describe the spectra of secondaries produced on both nucleon
\cite{KTM,KP,Sh5} and nuclear \cite{KTMS,Sh6} targets. In QGSM the 
high-energy interaction is determined by an exchange of one or several 
pomerons. One pomeron is coupled to a valence quark and diquark, and the 
others are coupled to sea quark-antiquark pairs. All elastic and inelastic 
processes are the result of a cut of different number of pomerons or a cut 
between pomerons \cite{AGK}. The inclusive spectra of secondaries are 
determined by convolutions of the quark and diquark momentum distributions 
and the functions describing their fragmentation into hadrons. All these 
functions are determined by their Regge asymptotics. When one nucleon 
interacts with several, the average number of pomerons is increased, 
which tends to soften the momentum distributions of all quarks and 
diquarks and, consequently, the secondary hadron spectra.

The functions $R^h_{A,B}(y)$ can be taken in the form \cite{Sh3,Sh4}   
\begin{equation}
R^h_A(y) = \frac{f^h(-y,<\nu >_{NA})}{<\nu >_{NA} f^h(-y,1)} \;,
\end{equation}
where $f^h(y,<\nu >)$ is the contribution from the interaction with $\nu$ 
nucleons to the spectrum of secondary particle $h$ in $NA$ collision.

The results of secondary spectra calculation in the QGSM depend on
several parameters. They can be fixed from $pp$ data. The values of 
parameters were changed slightly in comparison with \cite{Sh5}, because
new experimental data have appeared. The results for the $x_F$-spectra of 
secondary $\pi^+$, $\pi^-$, $K^+$ and $K^-$ are presented in Fig. 5. The 
agreement with the data \cite{Bre,AB} is quite reasonable. The description 
of secondary proton and antiproton spectra (agreement with the data is of 
the same order) one can find in \cite{ACKS}.   

The rapidity distributions of secondary $\pi^+$, $\pi^-$, $K^+$, $K^-$, 
$p$ and $\bar{p}$ measured in $Pb+Pb$ central collisions at 158 GeV/c per 
nucleon \cite{Sik} are compared with our calculations using Eq. (34) in 
Fig. 6 .In the case of pions the agreement is reasonable. The differences 
between the calculated curves and the data are larger than 10\%. In the 
case of secondary kaons we see that there is a problems. 

The spectrum of $K^-$ is in good agreement with the model, whereas for 
$K^+$ disagreement is about 20\%. In the case of change of the model 
parameters one will obtain the disagreement in Figs. 5c and 5d. Even more 
serious is the $K^+/K^-$ ratios. Experimentally this ratio at $y \sim 0$ 
is about 1.5, however in the model we can not obtain it more than 1.2. It 
means that the sea contribution (which leads to the same yields of $K^+$ 
and $K^-$) is too large in the model. It is possible, that the account for 
percolation/string fusion effect \cite{DDU,CKai,APS} as well as final 
state
interactions \cite{CS} can solve this problem. 

In the cases of $p$ and $\bar{p}$ the agreement with the data is good 
enough.

\section{Conclusions}

We presented some simplest and model independent formulae obtained with
assumption that high energy heavy ion collision can be consider as the 
superposition of independent nucleon-nucleon collisions. So any serious
disagreement of these predictions with the data can be considered as the 
signal for some collective interaction, say QGP formation. 

Contrary to the case of elementary particle physics the numerical 
results for heavy ion collisions should be obtained correspondently to
the experimental conditions. For example, if only events with secondary
production are registrated one should use Eqs.~(21) and (24), whereas if 
events with nuclear disintegration and without secondary production are 
also registrated, the cross sections $\sigma^{prod}_{AB}$, 
$\sigma^{prod}_{NB}$ and $\sigma^{inel}_{NN}$ should be replaced by
$\sigma^{inel}_{AB}$, $\sigma^{inel}_{NB}$ and $\sigma^{tot}_{NN}$,
respectively.

The results for rapidity spectra of secondaries are more model dependent.
It is possible that the effects of inclusive density saturations are
observed at CERN SppS energies, and the new RHIC data also give evidences 
for such effects.

\newpage

\begin{center}
{\bf Figure captions}\\
\end{center}

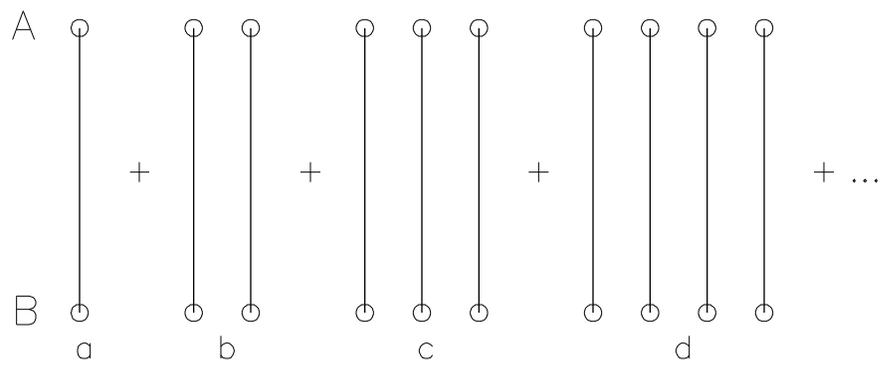
Fig. 1. The optical approximation diagrams for the scattering of two
nuclei.

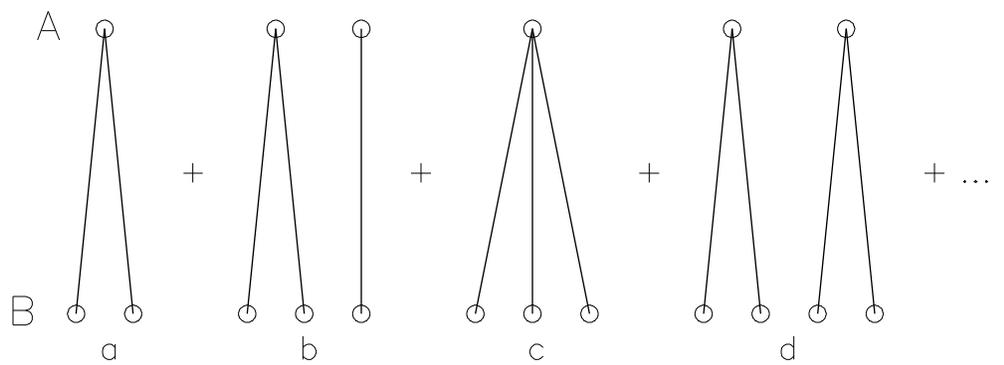
Fig. 2. The simplest corrections to the optical approximation which
correspond (together with the diagrams of Fig. 1) to the rigid target
approximation for the scattering of two nuclei.

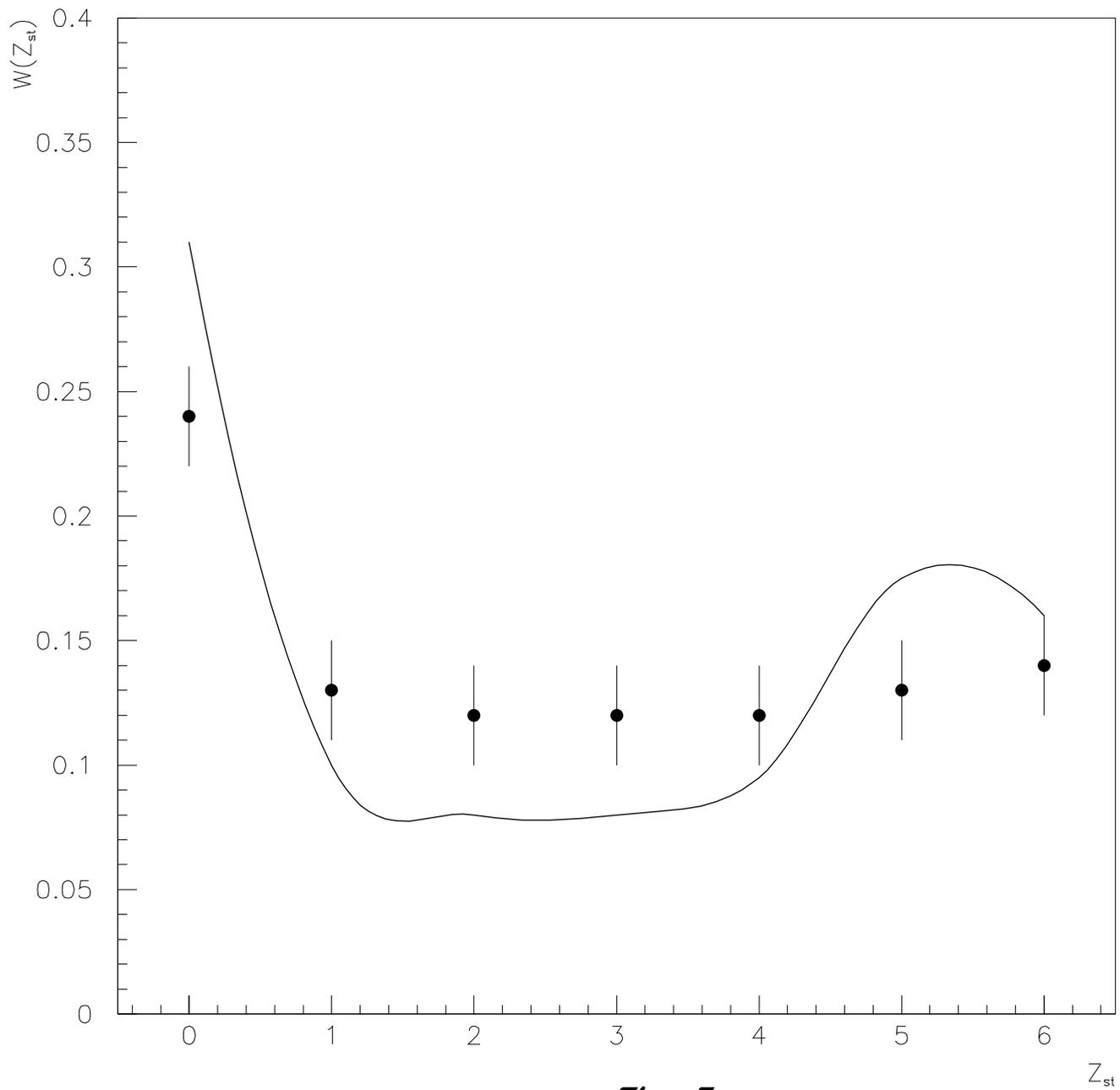
Fig. 3. The distribution on the stripping electrical charge of the
projectile in $C-Ta$ interactions at 4 GeV per nucleon and its
description by rigid target approximation \cite{CGS}.

Fig. 4. Predictions for the associated multiplicity distribution for
$W^{\pm}$ and $Z^0$ events (round black points) together with the 
experimental data ($W^{\pm}$ cross points and $Z^0$ white stars) and
the minimum bias distribution (squares). 

Fig. 5. The $x_F$-spectra of secondary $\pi^+$ (a), $\pi^-$ (b), $K^+$ (c) 
and $K^-$ (d), produced in $pp$ interactions at 100 and 175 GeV/c 
\cite{Bre} and 400 GeV/c \cite{AB}. The curves are the Quark-Gluon String 
Model predictions.  

Fig. 6. The rapidity spectra of secondary $\pi^+$, $\pi^-$ (a), $K^+$, 
$K^-$ (b) and $p$, $\bar{p}$ (c) produced in $Pb+Pb$ interactions at 158 
GeV/c per nucleon \cite{Sik}. The curves are the Quark-Gluon String Model
predictions.

\end {document}
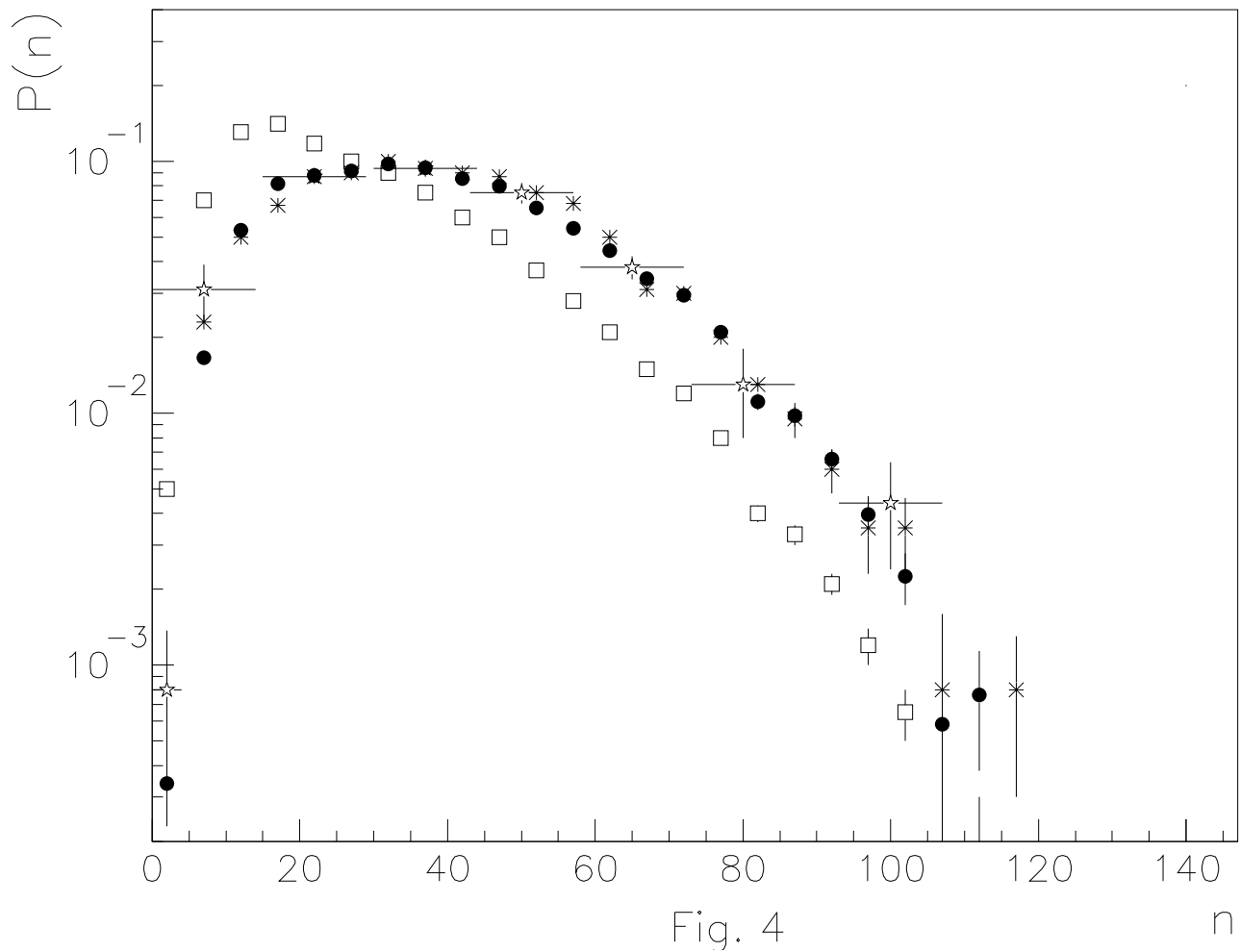
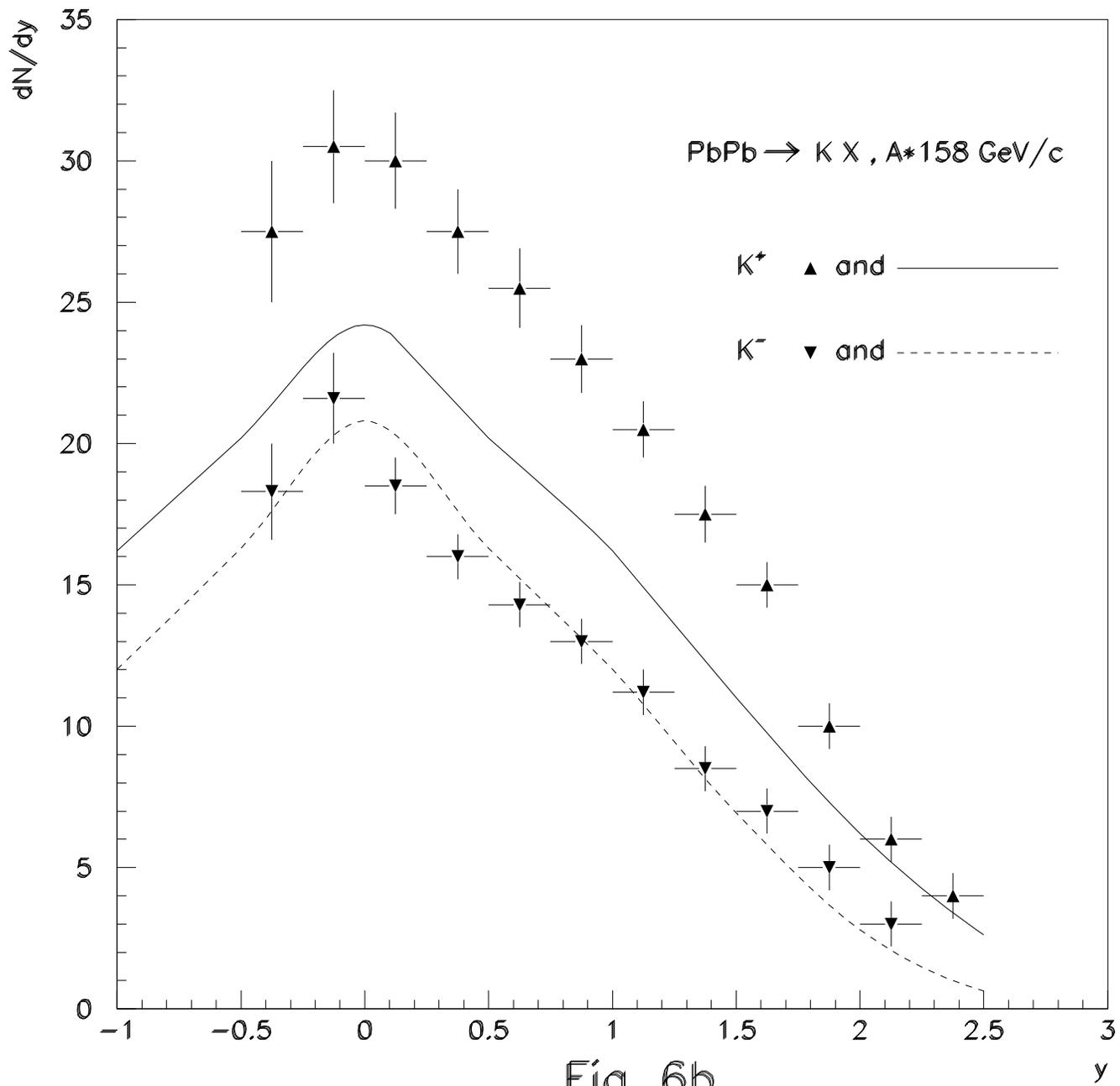
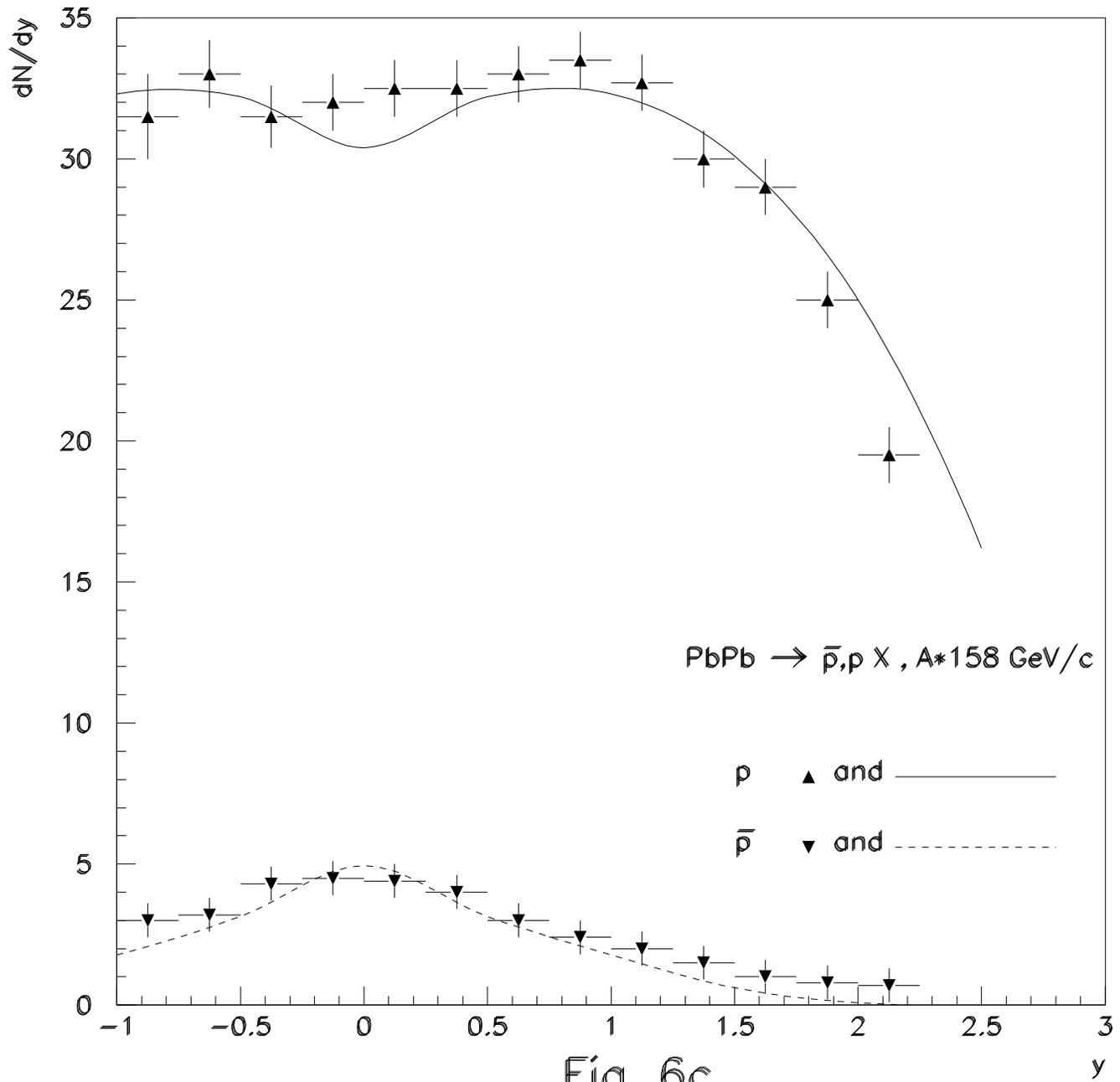
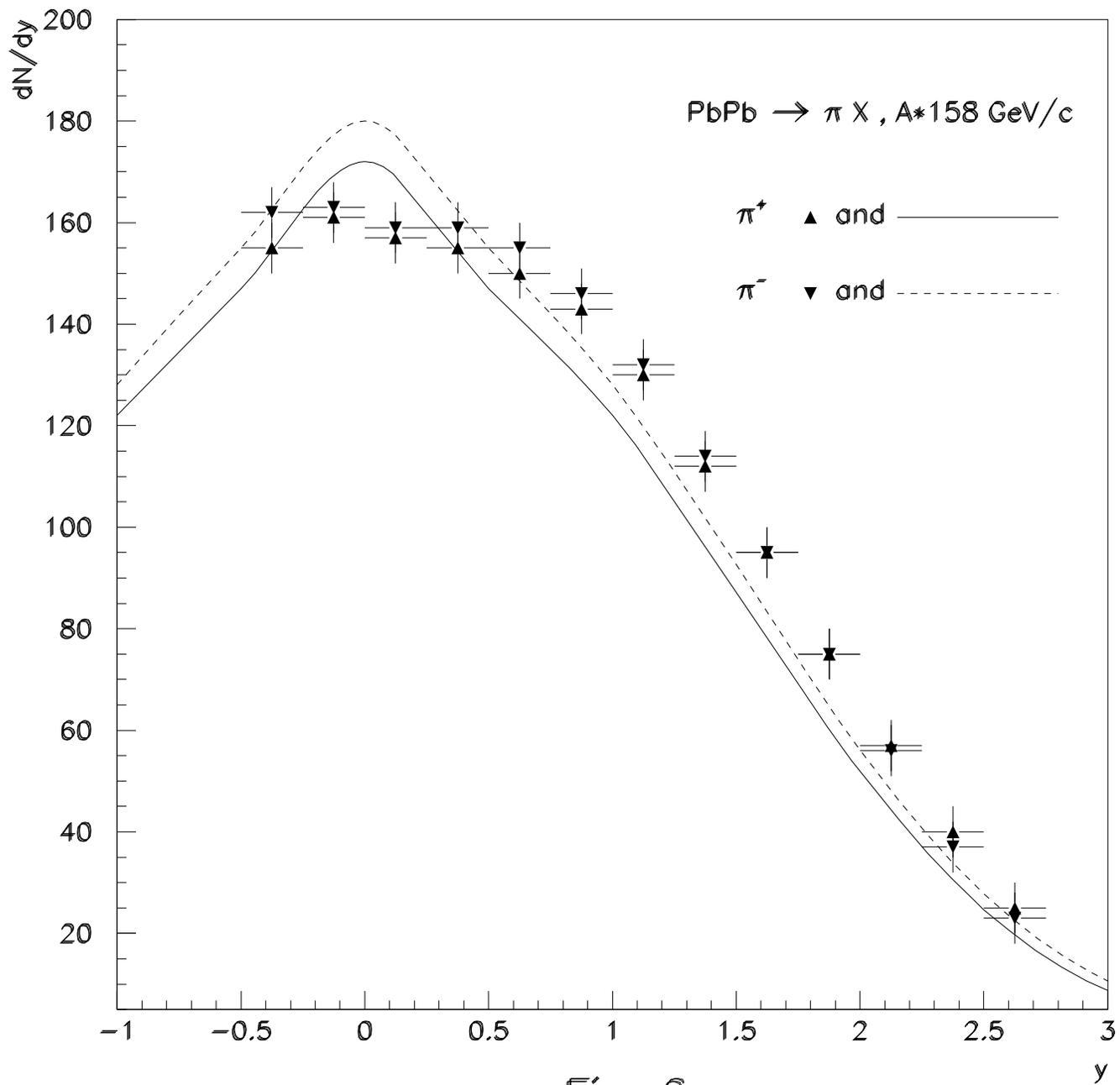
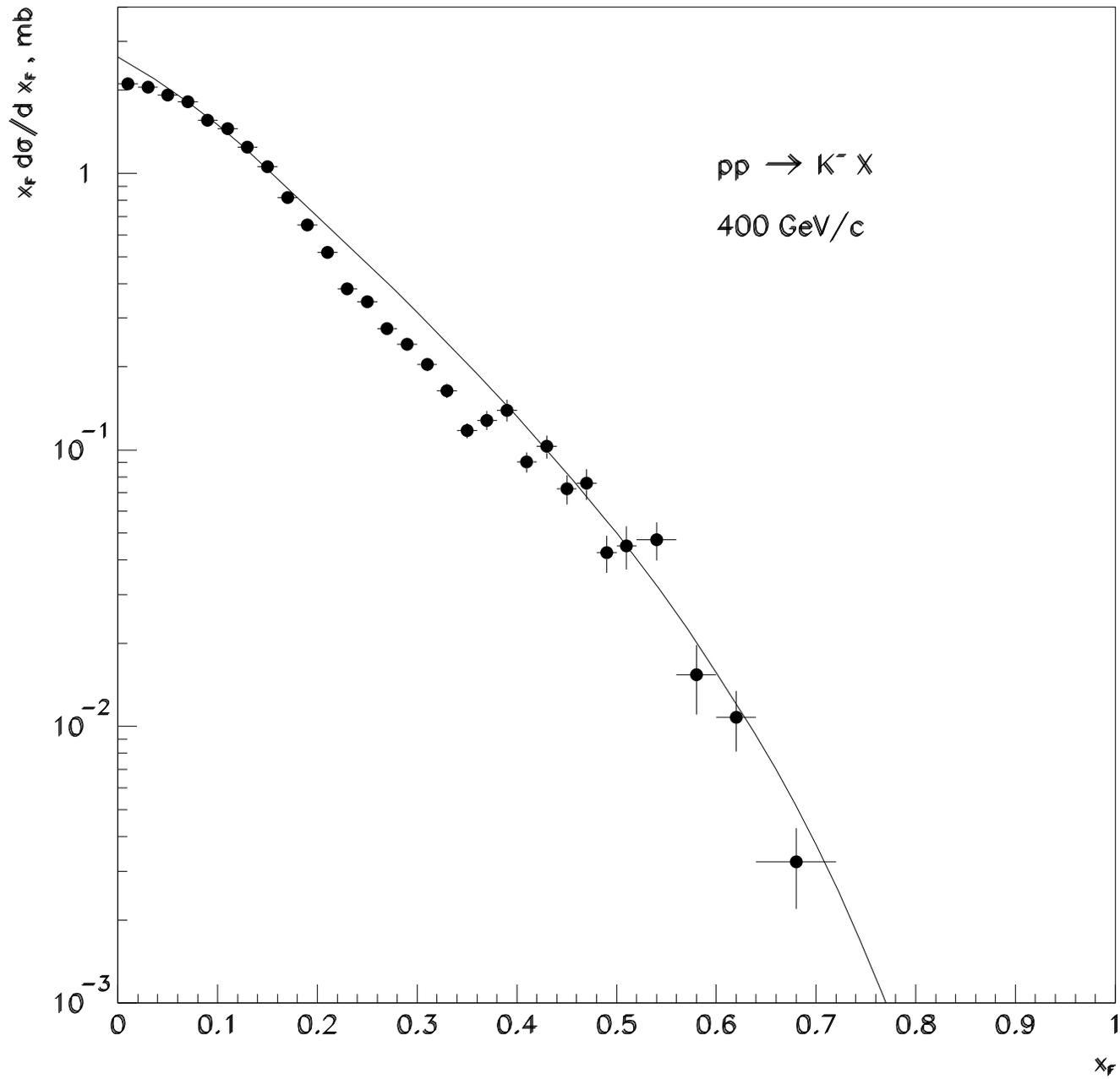
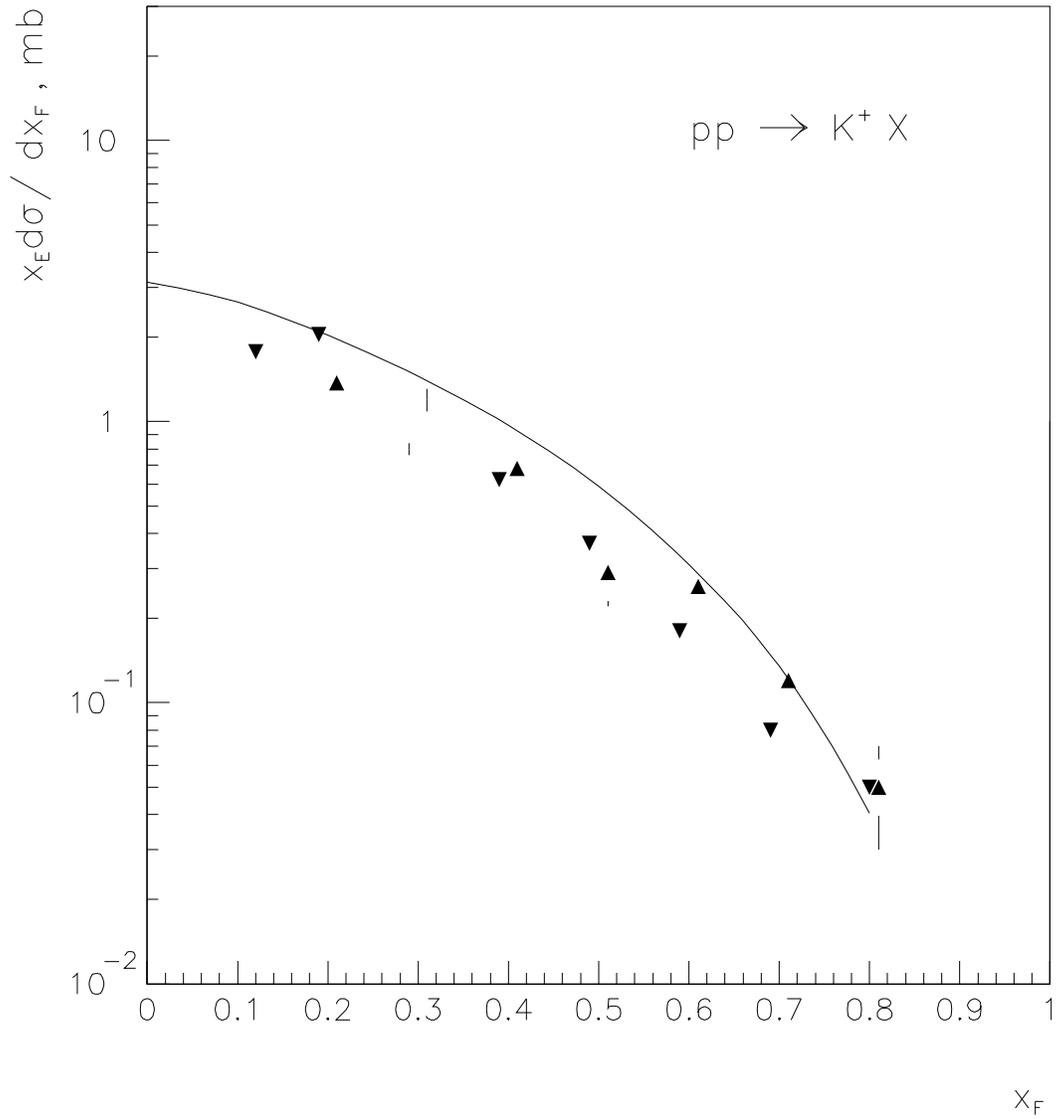
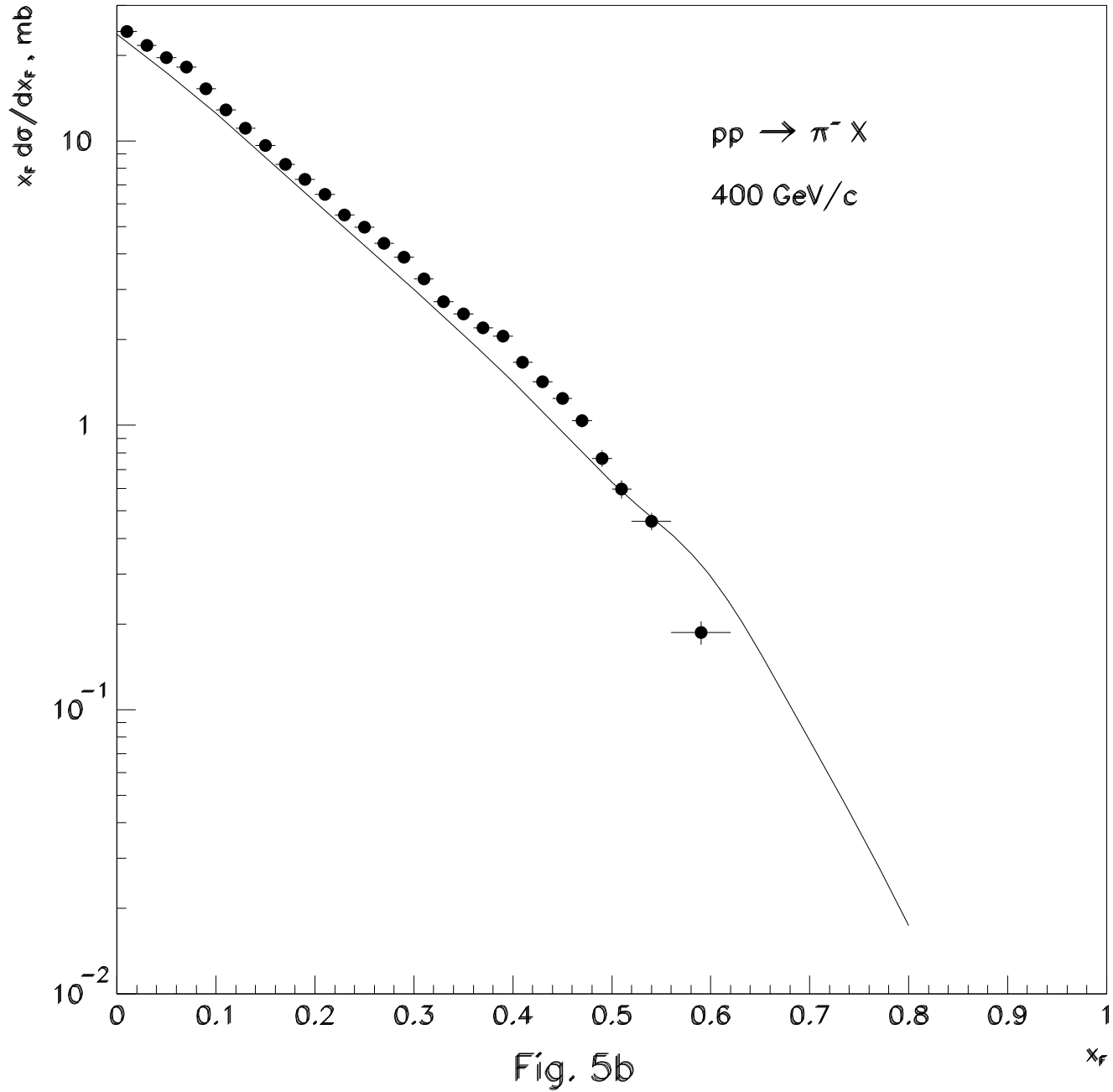
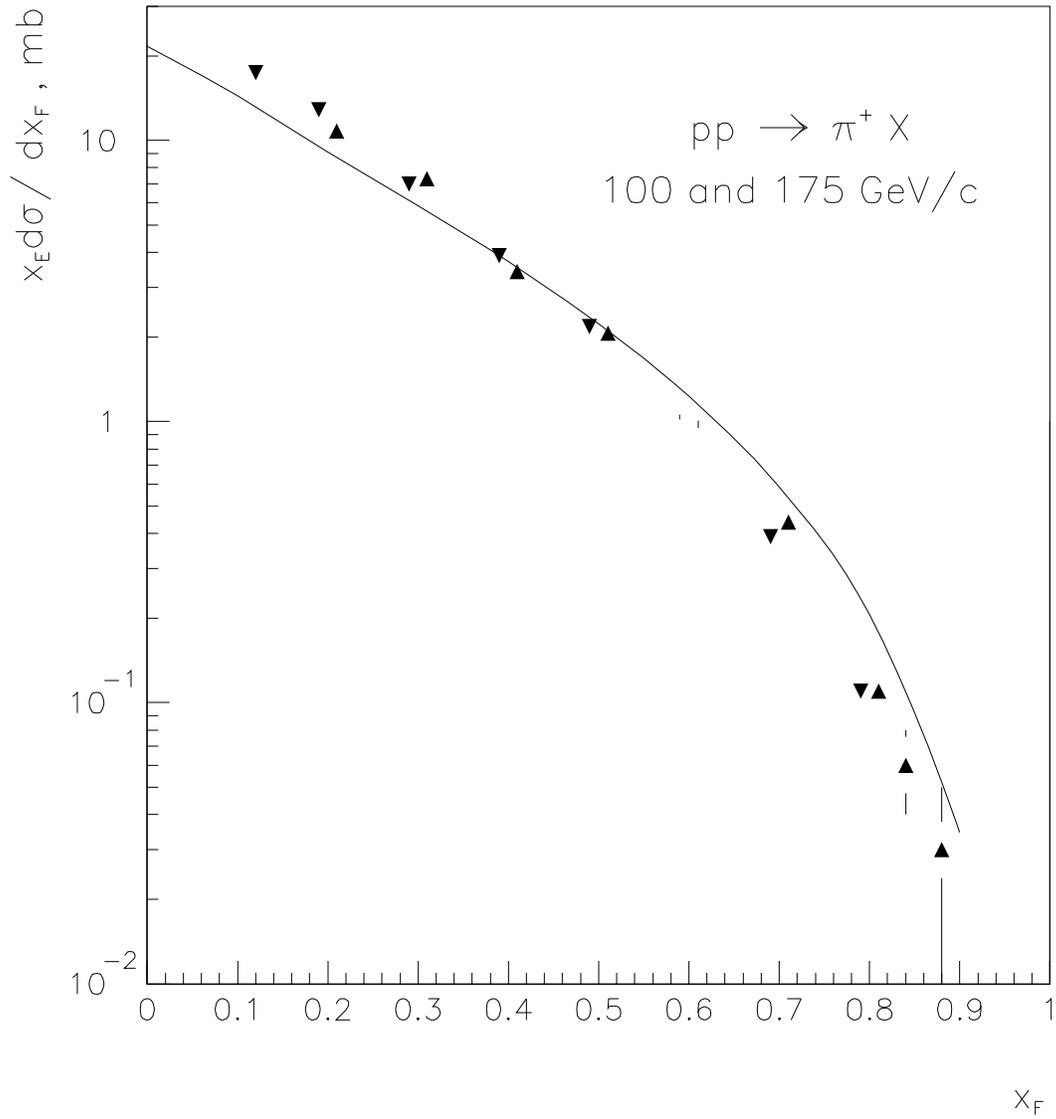
